\documentclass[english,aps,nofootinbib,longbibliography,notitlepage,superscriptaddress,prd,preprintnumbers,showkeys,twocolumn]{revtex4-1}
\pdfoutput=1

\usepackage[usenames,dvipsnames,svgnames,table]{xcolor}
\usepackage{verbatim}
\usepackage[T1]{fontenc}
\usepackage[latin9]{inputenc}
\usepackage{babel}
\usepackage{float}
\usepackage{graphicx}
\usepackage{svg}
\usepackage{hyperref}
\usepackage{amsthm,amsmath,amssymb,amsfonts}
\usepackage{todonotes}
\usepackage[mathscr]{eucal}
\usepackage{microtype}
\usepackage{esint}
\usepackage{amsbsy}
\usepackage{color}
\usepackage[capitalise]{cleveref}
\usepackage{breakurl}
\usepackage[normalem]{ulem}
\usepackage{bbm}
\usepackage{mathtools}
\usepackage{braket}
\usepackage{comment}
\usepackage[super]{nth}
\usepackage{csquotes}
\usepackage{lmodern}
\usepackage{todonotes}
\usepackage{hyphenat}

\definecolor{lightgray}{gray}{0.9}
\definecolor{Amber}{rgb}{1.0, 0.75, 0.0}
\definecolor{blizzardblue}{rgb}{0.67, 0.9, 0.93}

\newcommand{\dderiv}{\mathrm{d}}

\usepackage{tikz}
\usepackage{pgfplots}
\pgfplotsset{compat=newest,every axis plot/.append style={line width=1pt}}

\allowdisplaybreaks

\makeatletter

\pdfpageheight\paperheight
\pdfpagewidth\paperwidth

\@ifundefined{textcolor}{}{%
 \definecolor{BLACK}{gray}{0}
 \definecolor{WHITE}{gray}{1}
 \definecolor{RED}{rgb}{1,0,0}
 \definecolor{GREEN}{rgb}{0,1,0}
 \definecolor{BLUE}{rgb}{0,0,1}
 \definecolor{CYAN}{cmyk}{1,0,0,0}
 \definecolor{MAGENTA}{cmyk}{0,1,0,0}
 \definecolor{YELLOW}{cmyk}{0,0,1,0}
}

\makeatother

\makeatletter

\DeclareRobustCommand{\rcite}[1]{%
  \rcite@aux#1,\@nil{#1}%
}
\def\rcite@aux#1,#2\@nil#3{%
  \if\relax#2\relax
    Ref.~\cite{#3}%
  \else
    Refs.~\cite{#3}%
  \fi
}
\makeatother
\definecolor{wine}{RGB}{136,34,85}
\definecolor{teal}{RGB}{0,85,102}
\hypersetup{
    colorlinks = true,
    citecolor = {wine},
    linkcolor = {teal},
    urlcolor = {teal},
}

\newcommand{\be}{\begin{equation}}
\newcommand{\ee}{\end{equation}}
\newcommand{\ba}{\begin{eqnarray}}
\newcommand{\ea}{\end{eqnarray}}

\newcommand{\dd}{\mathrm{d}}

\newcommand{\lcdm}{$\Lambda$CDM }

\def\MPl{M_{\rm{Pl}}}

\def\half{\frac{1}{2}}

\def\phidot{\dot{\phi}}

\def\weff{w_{\rm{eff}}}

\def\Omm{\Omega_{\rm{m}}}
\def\OmM{\Omega_{\rm{m}}}

\def\rhoM{\rho_{\rm{m}}}
\def\PM{P_{\rm{m}}}
\def\HoverH{\frac{\dot{H}}{H^2}}

\begin{document}

\preprint{IFT-UAM/CSIC-24-83}

\title{To curve, or not to curve:\\ Is curvature-assisted quintessence observationally viable?}

\author{George Alestas}\email{g.alestas@csic.es}
\affiliation{Instituto de F\'isica Te\'orica (IFT) UAM-CSIC, C/ Nicol\'as Cabrera 13-15, Campus de Cantoblanco UAM, 28049 Madrid, Spain}
\author{Matilda Delgado}\email{matilda.delgado@uam.es}
\affiliation{Instituto de F\'isica Te\'orica (IFT) UAM-CSIC, C/ Nicol\'as Cabrera 13-15, Campus de Cantoblanco UAM, 28049 Madrid, Spain}
\author{Ignacio Ruiz}\email{ignacio.ruiz@csic.es}
\affiliation{Instituto de F\'isica Te\'orica (IFT) UAM-CSIC, C/ Nicol\'as Cabrera 13-15, Campus de Cantoblanco UAM, 28049 Madrid, Spain}
\affiliation{Departamento de F\'isica Te\'orica, Universidad Aut\'onoma
de Madrid, Cantoblanco, 28049 Madrid, Spain}
\author{\\Yashar Akrami}\email{yashar.akrami@csic.es}
\affiliation{Instituto de F\'isica Te\'orica (IFT) UAM-CSIC, C/ Nicol\'as Cabrera 13-15, Campus de Cantoblanco UAM, 28049 Madrid, Spain}
\affiliation{CERCA/ISO, Department of Physics, Case Western Reserve University, 10900 Euclid Avenue, Cleveland, Ohio 44106, USA}
\affiliation{Astrophysics Group and Imperial Centre for Inference and Cosmology, Department of Physics, Imperial College London, Blackett Laboratory, Prince Consort Road, London SW7 2AZ, United Kingdom}
\author{Miguel Montero}\email{miguel.montero@csic.es}
\affiliation{Instituto de F\'isica Te\'orica (IFT) UAM-CSIC, C/ Nicol\'as Cabrera 13-15, Campus de Cantoblanco UAM, 28049 Madrid, Spain}
\author{Savvas Nesseris}\email{savvas.nesseris@csic.es}
\affiliation{Instituto de F\'isica Te\'orica (IFT) UAM-CSIC, C/ Nicol\'as Cabrera 13-15, Campus de Cantoblanco UAM, 28049 Madrid, Spain}

\date{\today}

\begin{abstract}
Single-field models of accelerated expansion with nearly flat potentials, despite being able to provide observationally viable explanations for the early-time cosmic inflation and the late-time cosmic acceleration, are in strong tension with string theory evidence and the associated de Sitter swampland constraints. It has recently been argued that in an open universe, where the spatial curvature is negative (i.e., with $\Omega_k>0$), a new stable fixed point arises, which may lead to viable single-field-based accelerated expansion with an arbitrarily steep potential. Here, we show, through a dynamical systems analysis and a Bayesian statistical inference of cosmological parameters, that the additional cosmological solutions based on the new fixed point do not render steep-potential, single-field, accelerated expansion observationally viable. We mainly focus on quintessence models of dark energy, but we also argue that a similar conclusion can be drawn for cosmic inflation.
\end{abstract}

\preprint{}
\maketitle

\tableofcontents

\section{Introduction}
\label{sec:Introduction}
The Universe is accelerating \cite{SupernovaCosmologyProject:1998vns,SupernovaSearchTeam:1998fmf}, and it is believed that it underwent an accelerating expansion also at very early times \cite{Brout:1977ix,Starobinsky:1980te,Kazanas:1980tx,Sato:1981qmu,Guth:1980zm,Linde:1981mu,Albrecht:1982wi,Linde:1983gd,Planck:2018jri}. Since the observation of the late-time ``cosmic acceleration'' and the proposition that the Universe experienced an early-time phase of ``cosmic inflation,'' numerous attempts have been made to explain the two phenomena and to describe the mechanisms behind them based on fundamental theories; see, e.g., \rcite{Cicoli:2023opf} for a recent review. For the late-time acceleration, the theoretical efforts have been partially motivated by the notorious cosmological constant problem \cite{Weinberg:1988cp,Martin:2012bt}, but it has also been important, and interesting, to know whether a low-energy effective theory arising from some underlying ultraviolet (UV) completion can provide an explanation for the accelerating expansion.

We first provide, in this paper, an overview of the theoretical attempts in the context of string theory, discussing the de Sitter swampland conjecture and its regime of validity. We explain that it is difficult, if not impossible, to obtain models of (early- and late-time) accelerating expansion in the so-called asymptotic regions of moduli space where the theory is under perturbative control. Indeed, in all known examples ever checked, one obtains only effective single-field models with exponential scalar potentials that are way too steep to accommodate observationally viable acceleration. This theoretical ``evidence'' has convinced the majority of string theorists to focus their attention on the bulk of moduli space, beyond the perturbative regime, where computational control is much harder to attain.

In the rest of the paper, we investigate in detail the results of \rcite{Marconnet:2022fmx,Andriot:2023wvg}, where the authors argue that the aforementioned obstacles to finding a dynamical mechanism for cosmic acceleration and/or cosmic inflation in asymptotic regions of moduli space can be overcome if the Universe has a negative spatial curvature---the reason is the appearance of a fixed point in the phase space of the model that does not exist in the zero-curvature scenario. The arguments of \rcite{Marconnet:2022fmx,Andriot:2023wvg} are based on a phase space analysis of a scalar field model with an exponential potential and in the absence of matter and radiation. Here, we analyze the model through a more realistic setup and by comparing its cosmological predictions with observational data. We show, after a qualitative discussion of the phase space of the model and through both a numerical grid analysis and a full Bayesian statistical exploration of the parameter space of the model, that allowing the curvature parameter to take nonzero values does not help and the resulting ``curvature-assisted'' quintessence does not provide observationally viable cosmological solutions if the potential is steep. This is true even if we allow for the initial velocity of the scalar field to vary, in order to explore the full phase space. We place an upper bound on the slope of the scalar potential that is way too low for a single-field quintessence model to satisfy the theoretical de Sitter constraint. We also show, based on qualitative arguments, that the same conclusion holds for cosmic inflation.

The paper is organized as follows. In \cref{sec:ST-SF}, we discuss the importance of scalar fields in string theory and the theoretical constraints on their properties. We describe the difficulties in obtaining models of accelerating expansion in string theory and discuss possible ways to overcome the difficulties. In \cref{sec:curvature}, we focus on curvature-assisted quintessence and the question of whether a nonzero spatial curvature can help single-field models with steep potentials to provide solutions that are consistent with cosmological observations. We begin our investigation of the models in \cref{sec:Dyn_System_Anal} by reviewing the phase space and differential equations describing background cosmological solutions for an exponential potential and in the context of dynamical systems. In \cref{sec:quality}, we provide theoretical and mostly qualitative arguments for why including a nonzero curvature parameter does not make steep-potential, single-field models of dark energy and inflation observationally viable. Section \ref{sec:Cosmo} describes our cosmological exploration of the phase and parameter spaces of the model with an exponential potential, where we analyze the cosmological solutions in terms of their observational viability. We first perform, in \cref{sec:Num_Grid_Anal}, a numerical grid analysis of the phase space of the model by analyzing different solutions and their properties. We show that steep potentials do not provide, simultaneously, an extended epoch of matter domination and a sufficiently low value of the effective equation of state parameter at the present time, both of which are necessary conditions for a viable cosmological solution. Section \ref{sec:Bayes} presents the results of our full Bayesian statistical exploration of the parameter space of the model, where we constrain the parameters by confronting the model with a combination of recent cosmological data and demonstrate that the allowed upper bound on the slope of the scalar potential does not increase significantly when the spatial curvature takes nonzero values. We conclude in \cref{sec:Conclusions}.

\section{String theory and scalar fields}\label{sec:ST-SF}
A universe in accelerated expansion can be easily understood in effective field theories (EFTs) minimally coupled to gravity. The simplest way to do so is by means of a positive cosmological constant, $\Lambda>0$, or a slowly rolling scalar field in what is called a  ``quintessence'' scenario (see, e.g., \rcite{Copeland:2006wr}). With the aim of connecting our Universe to a UV-complete theory of quantum gravity such as string theory, one can ask which of these EFTs can be consistently embedded in a string theoretic setting. Indeed, after compactifying six of the ten spacetime dimensions and obtaining a four-dimensional macroscopic universe, string theoretic EFTs offer a rich phenomenology that can be used to reproduce the physics of our Universe.

In string theory,\footnote{This is expected for any consistent theory of quantum gravity \cite{McNamara:2020uza}.} there are no free parameters. For example, coupling constants in the EFT are not just numbers one chooses at will, but instead they are all dynamically controlled by the vacuum expectation values of some scalar fields $\vec\phi$. These fields are known as \emph{moduli}, and the scalar manifold over which they take values is known as \emph{moduli space}. There can also be a moduli-dependent potential $V(\vec\phi)$, which contributes to the total energy density of the Universe. A straightforward way to realize a positive $\Lambda$ would be to find a positive (de Sitter) minimum of this scalar potential, $\vec\phi_0$, resulting in an effective positive cosmological constant $\Lambda=8\pi G_\mathrm{N}\, V(\vec\phi_0)$.

There is a large body of string theory literature devoted to charting the moduli space and the associated potential $V(\vec\phi)$. The moduli space is often divided into the ``bulk'' region, where all moduli have a vacuum expectation value (VEV) of $\mathcal{O}(1)$, and the ``asymptotic'' regions, where one or more moduli have a very large VEV. Very little is known about the scalar potential in the bulk of moduli space, where there is no control parameter and the theory becomes subject to quantum corrections and nonperturbative effects. In contrast, in asymptotic limits of moduli space, oftentimes we can do a perturbative asymptotic expansion in some of the moduli $\phi^a$, and, consequently, $V(\vec\phi)$ can be reliably computed. Because of this, there has been a large effort to find de Sitter minima in the asymptotic regions of moduli space.  However, despite decades of effort, realizing a de Sitter minimum in these asymptotic regions remains an open question. In fact, every time we compute an asymptotic potential in string theory, it is a sum of exponentials:
\begin{equation} V(\vec{\phi})=V_0 e^{-\vec{\lambda}\cdot{\vec{\phi}}}\,,\label{expoform}\end{equation}
which, therefore, has no minima. This has been checked in many string constructions \cite{Grimm:2019ixq,McAllister:2023vgy,VanRiet:2023pnx}, for many moduli, and there are general arguments that this should be the case \cite{Castellano:2021mmx,Castellano:2022bvr}---there are certainly no counterexamples to \eqref{expoform}.

\begin{figure}
  \centering    \includegraphics[width=0.95\columnwidth]{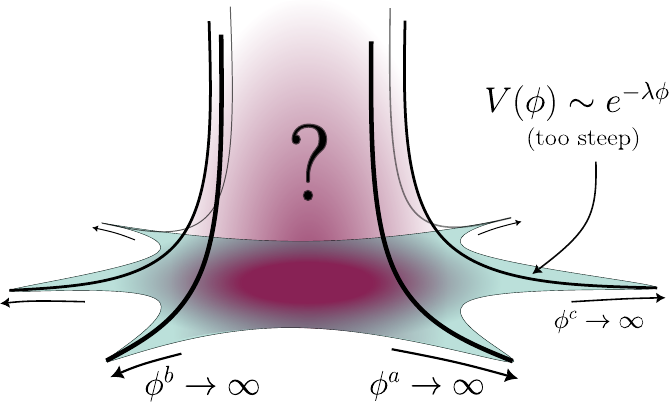}
    \caption{A schematic view of the moduli space of a string theoretic EFT. At the asymptotic boundaries of moduli space, where some field $\phi^a, \phi^b,\ldots$ receives a very large vacuum expectation value, the theory is under perturbative control, but we only ever obtain exponential potentials that are way too steep to accommodate acceleration. In the bulk of moduli space, the theory is strongly coupled and we lose computational control.}
    \label{figawesome}
\end{figure}

In fact, the situation is even worse. In all known asymptotic string theory examples, it is true that 
\begin{equation}\label{dSConj}
   \frac{\|\nabla V\|}{V}\geq \frac{2}{\sqrt{d-2}}\quad\text{as}\quad \vert\phi\vert\to\infty\,,
\end{equation}
where $d$ is the spacetime dimension and the norm is taken with respect to the metric in moduli space defined by the moduli kinetic fields, the $\mathsf{G}_{ab}$ in the action
\begin{equation}\label{action1-v1}
S=\int\dderiv^d x\sqrt{-g}\left\{\frac{1}{16\pi G_\mathrm{N}}R-\frac{1}{2}\mathsf{G}_{ab}\partial_\mu\phi^a\partial^\mu\phi^b-V(\vec\phi)\right\}.
\end{equation}
The gradient bound in \eqref{dSConj} implies that $V(\vec\phi)$ is way too steep for single-field models to accommodate eternally accelerating cosmological solutions (see also \rcite{Agrawal:2018own,Akrami:2018ylq,Raveri:2018ddi,Akrami:2020zfz} for observational bounds on the slope of the potential); therefore, not even a quintessence scenario can be realized in asymptotic corners of moduli space, at least in a naive way. The situation is summarized in \cref{figawesome}---there is a bulk moduli space that we do not understand very well, where in principle, quintessence or $\Lambda$ could lurk, and the asymptotic region, which we understand well, but does not accommodate accelerated expansion.

The statement \eqref{dSConj} is true in every example ever checked (see, for instance, \rcite{Maldacena:2000mw,Hertzberg:2007wc,Obied:2018sgi,Andriot:2019wrs,Andriot:2020lea,Calderon-Infante:2022nxb,Shiu:2023fhb,Shiu:2023nph,Cremonini:2023suw,Hebecker:2023qke,VanRiet:2023cca,Seo:2024fki}) and has been conjectured to hold universally \cite{Rudelius:2021azq} in what is known as the ``strong de Sitter conjecture'' in the ``swampland'' literature. Weaker versions of the conjecture, where the $2/\sqrt{d-2}$ on the right-hand side is replaced by other values \cite{Bedroya:2019snp} or just a generic $\mathcal{O}(1)$ number, have also been put forth, with different general arguments supporting them. We emphasize that all of these statements refer to the asymptotic regions of moduli space, a fact often misunderstood. 

In short, from a string theory point of view, it looks like we should leave the easy asymptotic region and start working our way toward the bulk (this is indeed what most proposed de Sitter constructions do; see \rcite{Kachru:2003aw,Danielsson:2012by,Blaback:2013ht,Kachru:2019dvo,Bena:2022cwb,Cicoli:2023opf} and references therein). However, the recent work \cite{Marconnet:2022fmx,Andriot:2023wvg} revived the interest in asymptotic regions in string theory, showing that one can get acceleration even when  $\lambda >\sqrt{2}$, simply by allowing the Universe to have a negative spatial curvature. This fact was already known in the cosmology literature \cite{vandenHoogen:1999qq,Gosenca:2015qha}, but \rcite{Marconnet:2022fmx,Andriot:2023wvg} emphasized that this would revive the hope that one could obtain accelerated expansion in a fully controlled string theoretic EFT. In the next section, we review this construction in some detail, pointing out a number of features that significantly challenge its cosmological viability.

\section{Curvature to the rescue?}\label{sec:curvature}
We now describe the dynamical system associated with a universe filled with a single-field quintessence as dark energy, following \rcite{vandenHoogen:1999qq,Gosenca:2015qha,Marconnet:2022fmx,Andriot:2023wvg}. We allow the spatial curvature to take a nonzero value, and we also include nonrelativistic matter in the analysis with the intention of comparing the system to observational constraints. Although in a fully realistic scenario radiation should also be included, we omit its contribution to the energy budget of the Universe as it becomes negligible long before the accelerated expansion phase begins. We use this approximate dynamical system to draw qualitative conclusions about the viability of quintessence dark energy, before constraining the model in the next section through a full statistical analysis of its parameter space with no approximation.

\subsection{Dynamical system}
\label{sec:Dyn_System_Anal}
\noindent

The action for a canonical scalar field $\phi$ minimally coupled to gravity and in the presence of nonrelativistic matter (composed of baryonic and cold dark matter) is given by
\begin{equation}
S = \int \dd^4 x \sqrt{-g} \left[ \frac{1}{2}\MPl^2 R - \frac{1}{2} g^{\mu \nu} \partial_{\mu}\phi \partial_{\nu}\phi -V(\phi) + \mathcal{L}_{\rm{m}} \right]\,,
\end{equation}
where $\MPl=1/\sqrt{8\pi G_\mathrm{N}}$ is the reduced Plank mass, and $\mathcal{L}_{\rm{m}}$ is the matter Lagrangian. We proceed by assuming the Friedmann-Lema\^{i}tre-Robertson-Walker metric
\begin{equation}
\dd s^2=-\dd t^2 + a(t)^2 \left( \frac{\dd r^2}{1-kr^2} + r^2 \dd\theta^2 +r^2\sin{\theta}^2 \dd\phi^2 \right),
\end{equation}
where $a(t)$ is the time-dependent scale factor and $k \in \{-1,0,+1\}$ is the spatial curvature constant, representing, respectively,
a homogeneous isotropic negatively curved (``hyperbolic'') universe,
a homogeneous isotropic flat (``Euclidean'') universe, and
a homogeneous isotropic positively curved (``spherical'') universe. The Friedmann equations are 
\begin{align}
3\MPl^2 H^2 &= \half \dot{\phi}^2 +V(\phi)+\rhoM - 3 \MPl^2 \frac{k}{a^2}\,,\label{eq:Friedmann:scalar:one}\\
2\MPl^2 \dot{H} &= - \dot{\phi}^2 - \rhoM +2 \MPl^2 \frac{k}{a^2}\,,\label{eq:Friedmann:scalar:two} 
\end{align}
where $H \equiv \dot{a}/a$ is the Hubble expansion rate and an overdot denotes a derivative with respect to cosmic time $t$. Note that we have assumed a pressureless matter component with the equation of state $w_\mathrm{m}=0$. The scalar field $\phi$ follows the equation of motion
\begin{equation}
\ddot{\phi} + 3H\dot{\phi} + \frac{\dd V(\phi)}{\dd\phi} =0\,.
\label{eq:continuity:scalar}
\end{equation}
Comparing \cref{eq:continuity:scalar} to the general continuity equation $\dot{\rho}+3H(\rho +P)=0$ for a perfect fluid with a symmetric stress-energy tensor $T^{\mu \nu}$ in a homogeneous and isotropic universe, the effective energy density and pressure of the scalar field take the forms
\begin{align}
\rho_\phi&=\dot{\phi}^2/2 +V(\phi)\,,\\
P_\phi&=\dot{\phi}^2/2 -V(\phi)\,,
\end{align}
respectively.
The equation of state parameter for the scalar field is then given by
\begin{equation}
w_{\phi}=\frac{P_{\phi}}{\rho_{\phi}}=\frac{\dot{\phi}^2/2 -V(\phi)}{\dot{\phi}^2/2 +V(\phi)}\,.
\end{equation}

To simplify the system of differential equations governing the background dynamics of the Universe and the evolution of the scalar field, we define the three dimensionless variables
\begin{equation}\label{eq:xyz_params}
x\equiv\frac{\phidot}{\sqrt{6} \MPl  H} \,, \quad y\equiv\frac{\sqrt{V(\phi)}}{\sqrt{3} \MPl  H}\,, \quad z\equiv-\Omega_k\equiv\frac{k}{a^2 H^2}\,,
\end{equation}
and
\begin{equation}\OmM\equiv\frac{\rho_\mathrm{m}}{3 \MPl ^2H^2}\,.\end{equation}
In terms of these variables, the first Friedmann equation \eqref{eq:Friedmann:scalar:one} becomes
\begin{equation}\label{eq:firstFried}
1=x^2+y^2-z+\OmM\,,
\end{equation}
and the ratio of the Friedmann equations
gives
\begin{equation}
\epsilon\equiv-\HoverH= \frac{3}{2} \left(x^2- y^2+1\right) + \frac{1}{2}z\,,
\end{equation}
where $\epsilon$ is the \emph{slow-roll parameter}. We also define the effective equation of state parameter
\begin{equation}
\weff \equiv\frac{P_{\phi}+\PM + P _z}{\rho_{\phi}+\rhoM + \rho_z}\,,\label{eq:w_eff}
\end{equation}
which can be written as
\begin{align}
\weff = \frac{2}{3}\epsilon-1= x^2 - y^2 + \frac{1}{3}z\,.\label{eq:w_eff_2}
\end{align}
For an exponential potential $V(\phi)=V_0e^{-\lambda \phi}$, the evolution of the dynamical system can now be described by a system of three coupled differential equations:
\begin{align}
   \frac{\dd x}{\dd N} &=\sqrt{\frac{3}{2}} \lambda  y^2-\frac{1}{2} x \left(-3 x^2+3 y^2-z+3\right)\,, \label{eq:dyn_system_1}\\
   \frac{\dd y}{\dd N} &=\frac{1}{2} y \left(3 x^2-\sqrt{6} \lambda  x-3 y^2+z+3\right)\,,\label{eq:dyn_system_2}\\
   \frac{\dd z}{\dd N} &=z \left(3 x^2-3 y^2+z+1\right)\,,\label{eq:dyn_system_3}
\end{align}
where we have used the number of $e$-folds $N\equiv\ln a$ as time variable.

Based on the results of \rcite{vandenHoogen:1999qq,Gosenca:2015qha}, it was emphasized in \rcite{Marconnet:2022fmx,Andriot:2023wvg} that solving these equations could lead to viable cosmological histories with accelerated expansion for $\lambda > \sqrt{2}$ when $k=-1$ (or $\Omega_k>0$). This sprung much excitement in the string theory community, as these are the kinds of exponential potentials that are easily encountered in string theoretic EFTs. The hope was that this could allow us to embed a realistic cosmology in the asymptotic limit of moduli space, where string theory is under perturbative control. Indeed, in \rcite{Marconnet:2022fmx,Andriot:2023wvg}, it was argued that one could obtain an infinite number of $e$-folds of accelerated expansion along some trajectories in phase space, with $\lambda > \sqrt{2}$. Although these statements are true, we now point out a couple of reasons why this scenario might not be as appealing as it appears at first sight.

\subsection{Quality of accelerating phases}\label{sec:quality}

\begin{figure*}[t!]
\includegraphics[width=0.655\textwidth]{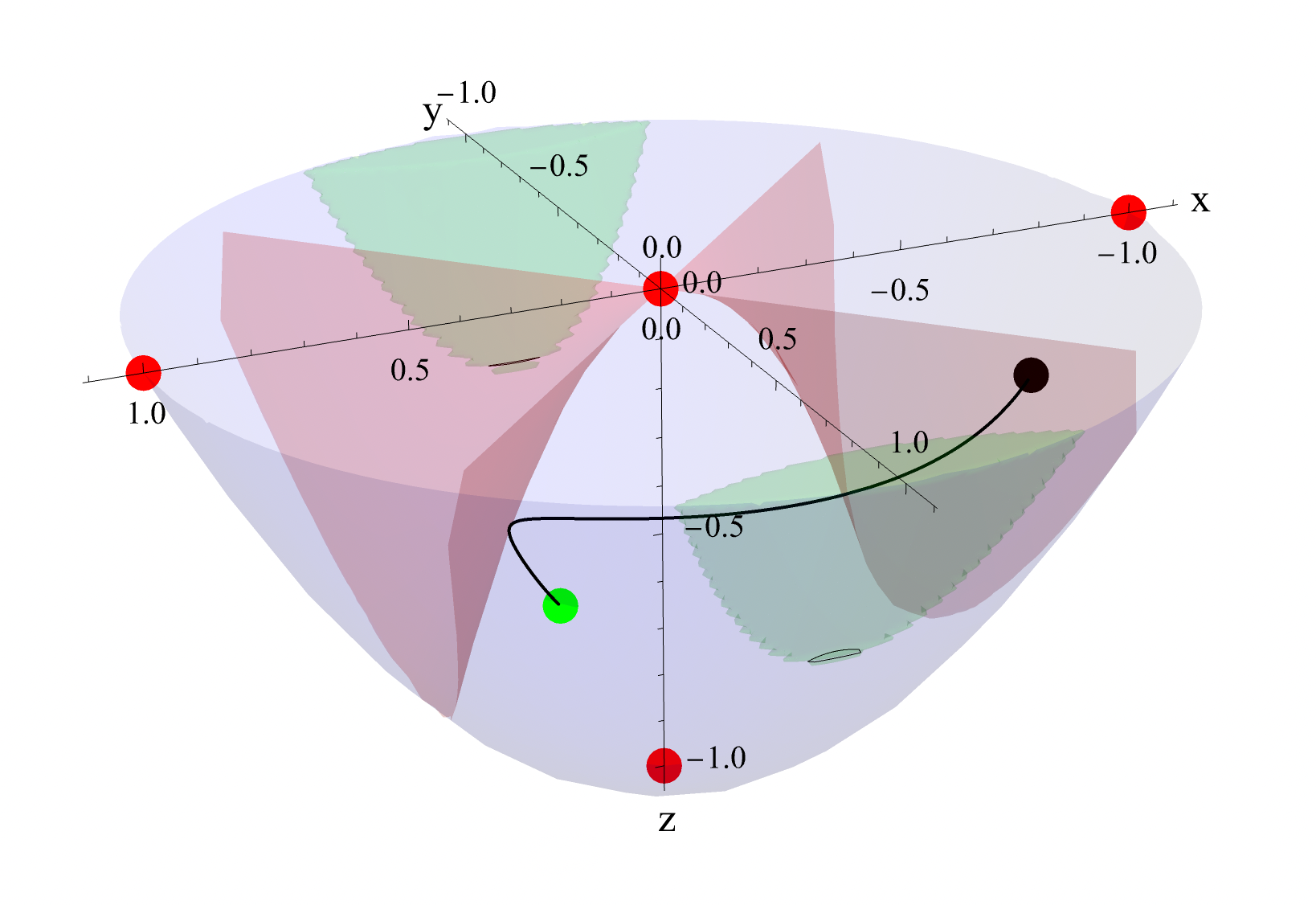}
\caption{\label{fig:phase_space} Phase space of the dynamical system described by \cref{eq:dyn_system_1,eq:dyn_system_2,eq:dyn_system_3} for a hyperbolic (or open) universe (i.e., for $k=-1$). The paraboloid corresponds to the Friedmann equation \eqref{eq:firstFried} with $\Omm=0$, above which all possible cosmic trajectories with nonzero $\Omm$ are located. Trajectories that pass through the green areas experience accelerating expansion with $\weff\in[-1,-0.7]$, while the red regions correspond to $\weff=0$ (i.e., matter domination); see \cref{eq:w_eff}. The red and green dots show the relevant fixed points of the system and correspond, respectively, to unstable and attractor solutions. For the example of $\lambda=1.6$, the attractor point behaves like a node, as shown by the indicative trajectory (black curve), where the black dot shows the initial conditions. } 
\end{figure*}

Figure \ref{fig:phase_space} shows the phase space of the dynamical system corresponding to \cref{eq:dyn_system_1,eq:dyn_system_2,eq:dyn_system_3} for a hyperbolic (or open) universe (i.e., for $k=-1$). The paraboloid in the figure corresponds to the Friedmann equation \eqref{eq:firstFried} with $\Omm=0$, above which all possible cosmic trajectories with nonzero $\Omm$ are located. The red and the green dots show the fixed points of the system that can potentially provide trajectories with desired properties and correspond, respectively, to unstable and attractor solutions. The figure also shows a trajectory for $\lambda=1.6$ (black curve) as an indicative example---the attractor point associated with the trajectory behaves like a node. For a more complete discussion of the fixed points and their stability see \rcite{Gosenca:2015qha,Marconnet:2022fmx,Andriot:2023wvg}.

Let us now choose polar coordinates on the phase space: $x= r \cos \theta$ and $y= r \sin \theta$. With these coordinates, the differential equations governing the dynamical system become
\begin{align}
\frac{\dd z}{\dd N}&= z\left(  1 + z +3 \cos(2\theta)r^2\right)\,, \label{rre34}\\
    \frac{\dd\theta}{\dd N}&= \frac{1}{2} (6 \cos\theta - \sqrt{6}\lambda r) \sin \theta\,,\\
    \frac{\dd r}{\dd N}&= \frac{r}{2}\left(  z +3  \cos(2\theta)(-1+r^2)\right)\,. 
\end{align}
Equation \eqref{rre34} can be rewritten as
\begin{equation}
\frac{\dd z}{\dd N}= z\,(3 \weff+1)\,,\label{e233}
\end{equation}
where we have used the expression \eqref{eq:w_eff_2}, $\weff= x^2 - y^2 + \frac{1}{3}z= \cos(2\theta)r^2+ \frac{1}{3}z$. Although in the previous subsection we assumed $w_\mathrm{m}=0$ and we ignored radiation, \cref{e233} holds even when these assumptions are dropped and $\weff$ is modified to include contributions of radiation and matter with pressure. 

Since accelerated expansion happens only when $\weff<-1/3$, \cref{e233} can be rearranged to provide an integral expression for the number of $e$-folds that an accelerated expansion epoch lasts:
\begin{equation} N=\int \dd N= \int_{\weff\leq -1/3}\frac{\dd z}{\vert z\vert (1+3\weff)}\,,\label{psch}\end{equation}
where the integral over $z$ is carried out only while $\weff\leq -1/3$. Since the allowed range of $z$ is finite ($-1 \leq z\leq 0$),  the number of $e$-folds that one can attain during the accelerating phase is finite away from the $z= 0$ and $1+3\weff=0$ loci. The locus $z=0$ is achieved only for the usual accelerated expansion present in a de Sitter or nearly flat quintessence potentials. Since $\Omega_k=-z$, this region corresponds to neglecting the effects of curvature, going back to the $k=0$ case, which has been thoroughly studied in the literature. For sufficiently steep potentials, i.e., for sufficiently large $\lambda$ (and in particular, for $\lambda>\sqrt{2}$), there is no fixed point with $z\approx0$, and any trajectories in that vicinity leave it quickly, generating only a small amount of $e$-folds. 

The key point in the curvature-assisted acceleration of \rcite{Marconnet:2022fmx,Andriot:2023wvg} is that the number of $e$-folds can also diverge when $1+3\weff$ approaches zero. As noted already in that reference, this is the boundary of the accelerating region---the trajectories found in \rcite{Marconnet:2022fmx,Andriot:2023wvg} with curvature asymptote to fixed points living in this boundary. Although the points themselves do not support accelerated expansion, a trajectory approaching it can support an arbitrarily large number of $e$-folds. For instance, for a fixed point at $z=z_*$, assuming $1+3\weff\sim C(z-z_*)$ in the vicinity of the fixed point, where $C$ is some constant, leads to $N\sim C^{-1}\log(\vert z-z_*\vert)$. 

Although, technically, we did obtain eternal accelerated expansion, the fact that we got it at the boundary of the region means that it is of very bad quality. In a precise sense, since $\weff$ is very close to $-1/3$, what we obtained is almost a decelerating cosmology, very different from the usual quintessence or de Sitter scenarios, which have $\weff\sim-1$. The amount of acceleration that an observer would measure is very small, and disappears very quickly---the slow-roll parameter
\begin{equation} \epsilon = \frac32(1+\weff)=1- (1+\epsilon_0)e^{-C N}\end{equation}
approaches 1 exponentially fast; after a few $e$-folds, this cosmology, for all intents and purposes, is indistinguishable from a nonaccelerating one. We will show this explicitly in the next section when we constrain the value of $\lambda$ with cosmological observations.

The curvature-assisted acceleration with steep potentials is also in serious trouble when describing an inflationary phase in the early Universe. For single-field, slow-roll cosmic inflation, there is an approximate relation \cite{Baumann:2018muz}
\begin{equation}r\sim 16\,\epsilon\end{equation}
between the tensor-to-scalar ratio $r$ and the slow-roll parameter $\epsilon$. $\epsilon\sim1$ (i.e., $\weff\sim-1/3$) has long been ruled out by observations; the current bound from the combination of {\it Planck} and BICEP2/Keck data is $r\lesssim0.04$ at $95\%$ CL \cite{Planck:2018jri,Tristram:2020wbi,BICEPKeck:2024stm}, which translates into $\epsilon\lesssim0.0025$. 

Furthermore, the fixed points with $\lambda > \sqrt{2}$ that lead to an arbitrary large number of $e$-folds of accelerated expansion with curvature in \rcite{Marconnet:2022fmx,Andriot:2023wvg} have $z_*=-\Omega_k$ of $\mathcal{O}(1)$ unless $\lambda$ is very close to $\sqrt{2}$. That means that for large values of $\lambda$, curvature is an $\mathcal{O}(1)$ part of the energy budget of the Universe during the eternally accelerated trajectories. This contrasts with observations, in either the inflationary or current era, which pose stringent upper bounds on $\Omega_k$.

All of the above show that the eternally accelerated trajectories in \rcite{Andriot:2023wvg} with $\lambda > \sqrt{2}$ are of no use for either inflation or to play the role of a quintessence field, since they are ruled out by observations. On the other hand, the current phase of accelerated expansion has lasted for only around $\sim 0.5$ $e$-folds, so one might entertain the possibility that a mild amount of curvature, compatible with observations,  may still drive a \emph{transient} phase of acceleration, even with a steep potential. Determining whether this is really the case can be done only through a careful study of the resulting cosmology, taking into account all observations. The remainder of this paper is devoted to this question.

\section{Cosmological constraints}
\label{sec:Cosmo}
\subsection{Numerical grid analysis of phase space}
\label{sec:Num_Grid_Anal}

Before we confront the curvature-assisted quintessence model with cosmological data, we briefly study the properties of its late-time cosmological solutions through a numerical grid analysis of the dynamical system described by \cref{eq:dyn_system_1,eq:dyn_system_2,eq:dyn_system_3} and presented in \cref{fig:phase_space}. This will further demonstrate that large values of $\lambda$ (i.e., $\lambda>\sqrt{2}$) are incapable of providing observationally viable cosmic histories even with nonzero curvature. Broadly speaking, there are two conditions that any viable solutions must satisfy: (i) provide an extended matter domination phase before the onset of cosmic acceleration, and (ii) yield an effective equation of state parameter $\weff$, that is $\sim-0.7$ at the present time, as for the best-fit $\Lambda$CDM model.

\begin{figure}
\centering
\includegraphics[width=0.41\textwidth]{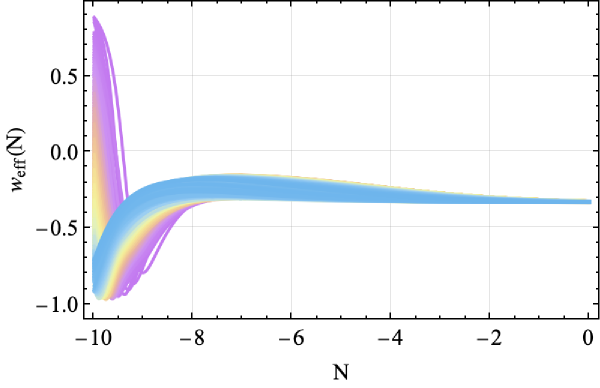}
\caption{\label{fig:w_eff} Evolution of $\weff$ as a function of the number of $e$-folds $N$ for quintessence with $\lambda=1.6$ and for trajectories that provide an epoch of $\weff\leq-0.7$, i.e., for those which go through the green regions in \cref{fig:phase_space}---these trajectories exhibit late-time accelerated expansion consistent with cosmological observations but do not contain a matter domination epoch.}
\end{figure}

Here, we solve the system of differential equations \eqref{eq:dyn_system_1}-\eqref{eq:dyn_system_3} for different initial conditions, i.e., for different initial values of the three phase space variables $\{x,y,z\}$, and then study the behavior of the cosmological background solution in each case. We do this through a fine grid scan of the entire phase space contained within the paraboloid in \cref{fig:phase_space}, defined as $x^2+y^2-z\leq1$; see \cref{eq:firstFried}. We compute all the trajectories for the different sets of initial conditions and keep only the ones that provide an epoch of $\weff\leq-0.7$, i.e., solutions which go through the green regions in \cref{fig:phase_space}---we show one example of such solutions for $\lambda = 1.6$ in the figure (black curve). Figure \ref{fig:w_eff} presents the time evolution of $\weff$ for all these accelerating solutions. It is clear from the figure that none of the solutions exhibit an epoch of matter domination in the past, i.e., before the accelerating phase of cosmic history. In other words, none of the solutions stay within the red regions in \cref{fig:phase_space} for a sufficiently long time.

\begin{figure}
  \hspace*{-0.4cm} 
  \includegraphics[width=0.47\textwidth]{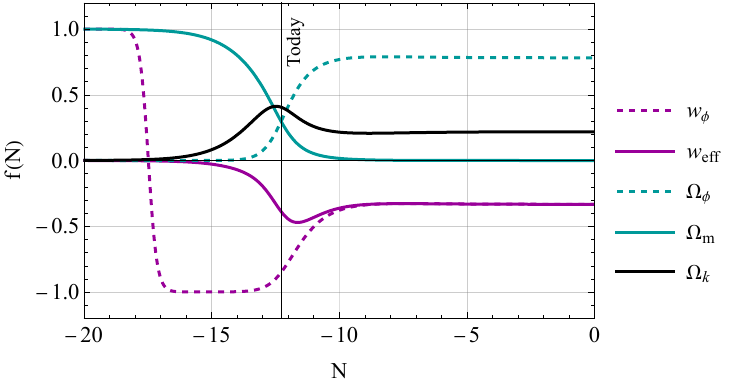}
  \includegraphics[width=0.49\textwidth]{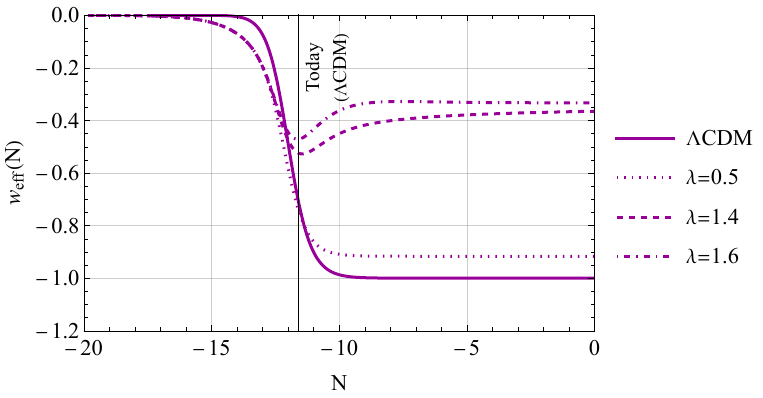}
  \caption{\label{fig:evol_lambda}{\bf Upper panel:} evolution of $w_{\phi}$ (equation of state parameter for the scalar field $\phi$), $\weff$ (effective equation of state parameter), $\Omega_{\phi}$ (scalar field density parameter), $\Omm$ (nonrelativistic matter density parameter), and $\Omega_{k}$ (curvature density parameter) as functions of the number of $e$-folds $N$ for $\lambda = 1.6$. The vertical line corresponds to the present time at which $\Omm\sim 0.3$. {\bf Lower panel:} $\weff(N)$ for quintessence with $\lambda = 0.5$, $\lambda = 1.4$, and $\lambda = 1.6$, in comparison with $\weff(N)$ for a good-fit \lcdm model with $\Omega_k$ set to zero. The vertical line corresponds to the present time for the \lcdm example.}
\end{figure}

Having said that, we do find solutions that provide extended matter domination periods, as we have demonstrated in the upper panel in \cref{fig:evol_lambda} for $\lambda=1.6$ as an example---see how $\Omm$ behaves with time (or number of $e$-folds). However, the problem with these solutions is that none of them provide an effective equation of state parameter $\weff$ less than $\sim-0.5$ at the present time (i.e., when $\Omm\sim 0.3$). This means that we do not obtain sufficient acceleration in these cases where matter domination exists. In the lower panel in \cref{fig:evol_lambda}, we present three examples of $\weff$ evolution for the curvature-assisted quintessence model, i.e., for $\lambda=\{0.5, 1.4,1.6\}$, and compare them with that of a \emph{flat} \lcdm model (i.e., with $\Omega_k$ set to zero) that provides a good fit to the data. The figure shows that by increasing $\lambda$ it becomes more and more difficult for the quintessence model to provide a cosmic history that is compatible with cosmological observations, as the current value of $\weff$ does not reach the observationally preferred value of $\sim-0.7$, as indicated by our \lcdm example.

In summary, the phase space of the quintessence model either does not contain solutions with extended matter domination or does not provide sufficiently large cosmic acceleration.

\subsection{Bayesian analysis and parameter estimation}
\label{sec:Bayes}

We now perform a full Bayesian statistical analysis of curvature-assisted, single-field quintessence to scan the entire parameter space of the model and constrain, particularly, the $\lambda$ parameter. This will tell us whether large values of $\lambda$ (i.e., $\lambda>\sqrt{2}$) preferred by the statement \eqref{dSConj} become observationally viable when the spatial curvature of the Universe is nonzero. We have already argued, based on qualitative discussions, as well as through a numerical grid analysis of the phase space of the model, that it is not the case, but here, we demonstrate that through a detailed comparison of the predictions of the model with some of the most recent cosmological observations. We employ the cosmic microwave background (CMB) distance priors provided by {\it Planck} \cite{Efstathiou:1998xx,Nesseris:2006er,Planck:2018vyg}, a baryon acoustic oscillation (BAO) data compilation \cite{Alestas:2022gcg}, and the Pantheon$+$ type Ia supernovae (SNe Ia) data \cite{Riess:2021jrx}.\footnote{We do not use the recent constraints provided by the Dark Energy Spectroscopic Instrument (DESI) \cite{DESI:2024mwx}, in order to avoid potential systematic effects, as the official DESI likelihood is not yet publicly available. In any case, pre-DESI cosmological data are sufficient for the purpose of the present paper, which is to show that large values of the parameter $\lambda$ are excluded observationally even if the Universe is negatively curved---the DESI results would not increase the upper bound on $\lambda$ that we report here.} The distance prior parameters are comprised of the acoustic scale $I_\mathrm{A}$, which measures the temperature of the CMB in the transverse direction, the shift parameter $R$, which measures the peak spacing of the temperature in the power spectrum, and the baryon density $\omega_\mathrm{b}\equiv\Omega_\mathrm{b}h^{2}$, where $\Omega_\mathrm{b}$ is the baryon density parameter and $h\equiv H_0/100$ with $H_0$ the present value of the Hubble expansion rate $H$. As for the BAO compilation, we use the data from WiggleZ \cite{Blake:2012pj, Escamilla-Rivera:2016qwv}, 6dFGS \cite{Beutler:2011hx}, the Dark Energy Survey \cite{DES:2021esc}, and the fourth generation of the Sloan Digital Sky Survey \cite{eBOSS:2020yzd,duMasdesBourboux:2020pck}.

We perform a Markov chain Monte Carlo (MCMC) scan of the parameter space of the quintessence model for two cases.\footnote{The {\it Mathematica} codes used in our analysis will be made publicly available in a GitHub repository upon publication of the paper.} We first consider a flat universe; i.e., we fix the present value of the curvature density parameter $\Omega_{k,0}$ to zero. We then allow $\Omega_{k,0}$ to vary as a free parameter. For each case, we perform two separate analyses, one by setting the initial value of the scalar field's time derivative $\dot{\phi}_0$ to zero and one by varying it as a free parameter. In both cases, we fix $\phi_0$ (the initial value of the scalar field) to zero and vary the parameter $V_0$ in the potential $V(\phi)=V_0e^{-\lambda \phi}$, as one of the two parameters is redundant and can be replaced by the other. We impose flat (or uniform) priors on all free parameters---\cref{tab:MCMC_priors} shows the ranges of parameters we employ in our scans. Comparing the parameter constraints we obtain for different cases will tell us whether assisting the accelerated expansion with a nonzero spatial curvature will increase the upper bound on the parameter $\lambda$ so that large values suggested by the statement \eqref{dSConj} become observationally viable.

\begin{table}
\rowcolors{1}{}{lightgray}
    \centering
    \setlength\tabcolsep{0pt}
    \begin{tabular}{ |c|c| }
    \hline
    \rowcolor{blizzardblue}
    ~Parameter~ & ~Prior range~ \\
    \hline
    $\Omega_{\rm{m,0}}$ & $[0.001, 1]$\\
    $\omega_\mathrm{b,0}$ & $[0.01, 0.04]$\\
    $\Omega_{k,0}$ & $[-0.2, 0.2]$\\
    $V_0$ & $[0, 3]$\\
    $\lambda$ & $[0, 3]$\\
    $h$ & $[0.5, 1]$\\
    $\dot{\phi}_0$ & $[-1, 1]$\\
    \hline
    \end{tabular}
    \caption{Ranges of model parameters employed in our scans. Note that $\dot{\phi}_0$ is fixed to zero in some of the analyses, so the presented range for $\dot{\phi}_0$ is only for cases where it is varied as a free parameter. $\Omega_{\rm{m,0}}$, $\omega_\mathrm{b,0}$, and $\Omega_{k,0}$ are the present values of $\Omega_{\rm{m}}$, $\omega_\mathrm{b}$, and $\Omega_{k}$, respectively.}\label{tab:MCMC_priors}
\end{table}

\begin{figure*}[t!]
\includegraphics[width=1\textwidth]{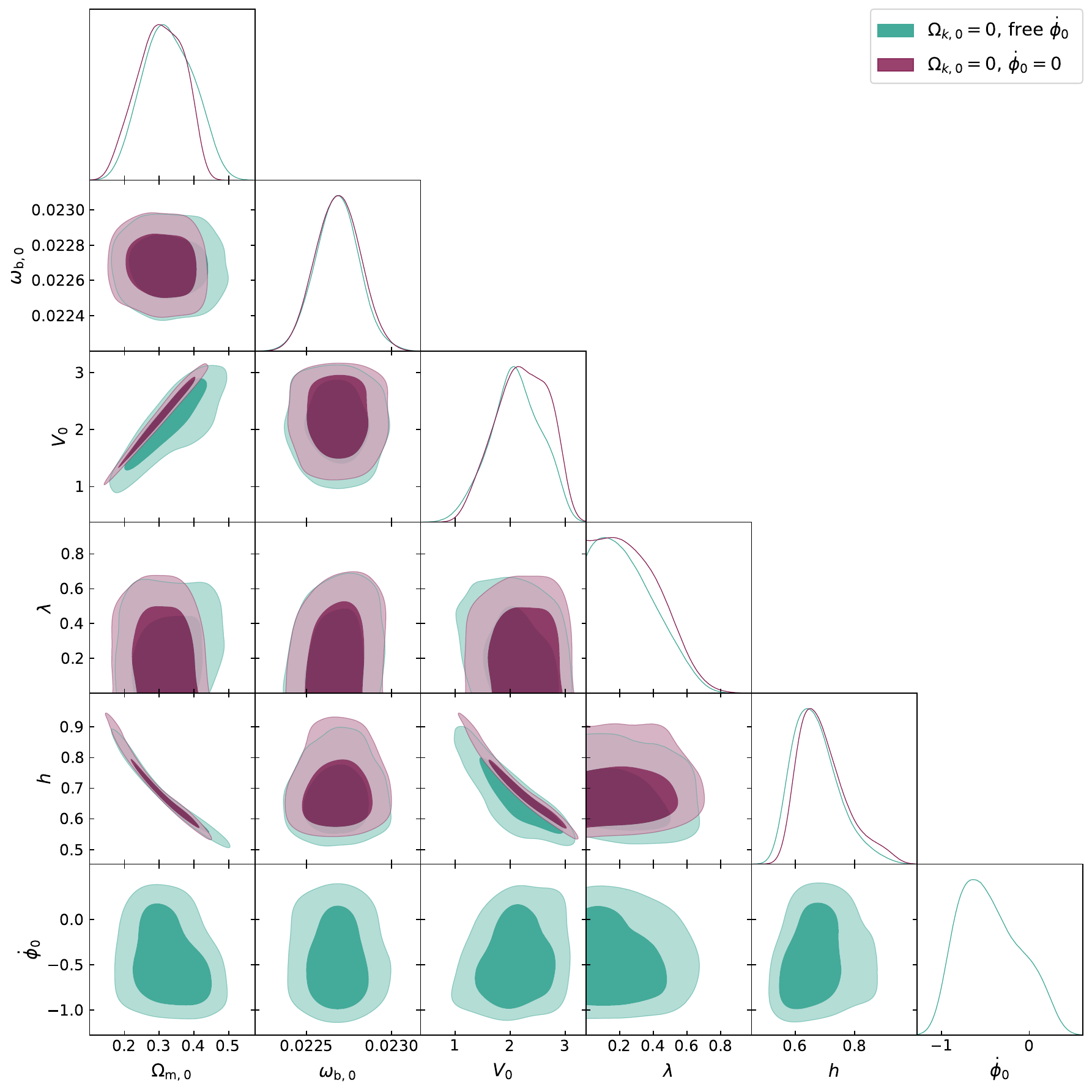}
\caption{\label{fig:mcmc_flat}
The $68.3\%$ and $95.5\%$ confidence regions and one-dimensional marginalized probability density functions for the quintessence model of dark energy with an exponential potential when the spatial curvature of the Universe is set to zero (i.e., $\Omega_{k,0} = 0$, where $\Omega_{k,0}$ is the present value of the curvature density parameter $\Omega_k$). We show the results for two cases: (i) the initial value of the scalar field's time derivative $\dot{\phi}_0$ is set to zero, and (ii) $\dot{\phi}_0$  is varied as a free parameter.}
\end{figure*}

\begin{figure*}[t!]
\includegraphics[width=1\textwidth]{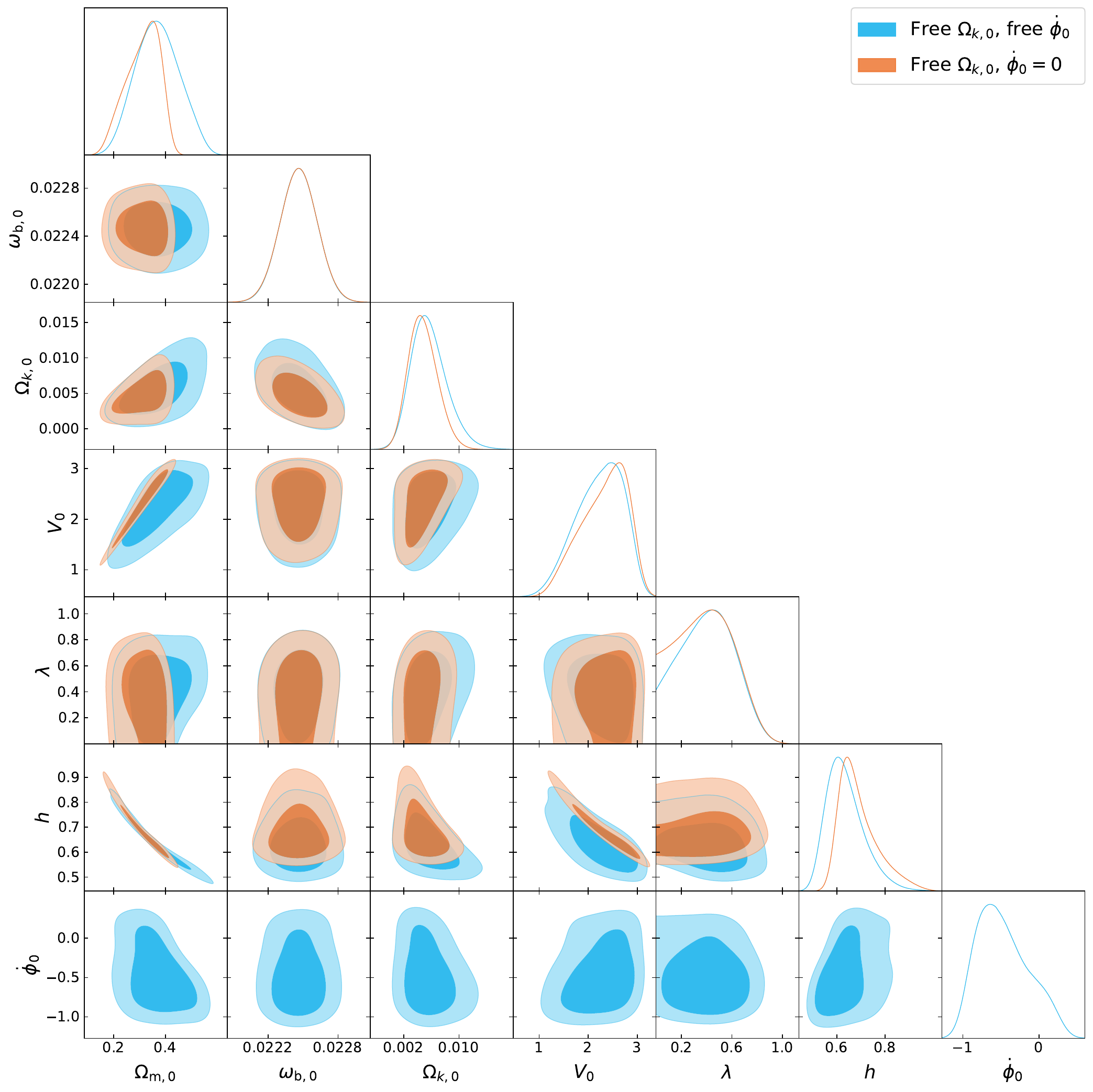}
\caption{\label{fig:mcmc_ok}
As in \cref{fig:mcmc_flat}, but for cases where the spatial curvature of the Universe is allowed to be nonzero; i.e., the present value of the curvature density parameter, $\Omega_{k,0}$, is varied as a free parameter.}
\end{figure*}

\begin{table*}
\renewcommand{\arraystretch}{1.3}
\rowcolors{1}{}{lightgray}
    \centering
    \setlength\tabcolsep{0pt}
    \begin{tabular}{ |c|c|c|c|c| }
    \hline
    \rowcolor{blizzardblue}
    ~Parameter~ & ~~$\Omega_{k,0}=0$ (flat universe), $\dot{\phi}_0 = 0$~~ & ~~$\Omega_{k,0}=0$ (flat universe), free $\dot{\phi}_0$~~ & ~~Free $\Omega_{k,0}$, $\dot{\phi}_0 = 0$~~ & ~~Free $\Omega_{k,0}$, free $\dot{\phi}_0$~~ \\
    \hline
    $\Omega_{\rm{m,0}}$ & $0.3050^{+0.0784}_{-0.0591}$ & $0.3269^{+0.0730}_{-0.0777}$ & $0.3134^{+0.0849}_{-0.0441}$ & $0.3699^{+0.0820}_{-0.0900}$ \\
    $\omega_\mathrm{b,0}$ & $0.0226^{+0.0002}_{-0.0001}$ & $0.0226^{+0.0002}_{-0.0001}$ & $0.0225^{+0.0001}_{-0.0002}$ & $0.0224^{+0.0002}_{-0.0001}$ \\
    $\Omega_{k,0}$ & $\cdots$ & $\cdots$ & $0.0047^{+0.0018}_{-0.0023}$ & $0.0056^{+0.002}_{-0.003}$ \\
    $h$ & $0.6911^{+0.0397}_{-0.1026}$ & $0.6684^{+0.0552}_{-0.0944}$ & $0.6870^{+0.0290}_{-0.0962}$ & $0.6345^{+0.0498}_{-0.0928}$ \\
    $\lambda$ & $\lesssim0.6$ & $\lesssim0.5$ & $\lesssim0.7$ & $\lesssim0.7$ \\
    $V_0$ & $2.2233^{+0.6312}_{-0.3597}$ & $2.1066^{+0.5234}_{-0.4686}$ & $2.3258^{+0.6579}_{-0.2345}$ & $2.2384^{+0.6710}_{-0.3018}$ \\
    $\dot{\phi}_0$ & $\cdots$ & $-0.4507^{+0.2021}_{-0.5001}$ & $\cdots$ & $-0.4045^{+0.1670}_{-0.4946}$ \\
    \hline
    \end{tabular}
    \caption{The $1\sigma$ (i.e., $68.3\%$ confidence level) constraints on model parameters obtained through MCMC scans of the parameter space for different cases.}
    \label{tab:MCMC_bf}
\end{table*}

We present the results of our MCMC analysis in \cref{fig:mcmc_flat,fig:mcmc_ok}, where two-dimensional $68.3 \%$ and $95.5 \%$ confidence regions and one-dimensional marginalized probability density functions (PDFs) for different parameters and different cases are provided. We also present in \cref{tab:MCMC_bf} $1\sigma$ (i.e., $68.3\%$ confidence level) constraints on the parameters. Since the parameter of interest here is $\lambda$, let us compare our constraints on that parameter as given in the figures and the table for the different cases we have studied. Figure~\ref{fig:mcmc_flat} and \cref{tab:MCMC_bf} show that cosmological observations place an upper bound of $\sim 0.6$ on $\lambda$ when $\dot{\phi}_0$ is fixed to zero and an upper bound of $\sim0.5$ when $\dot{\phi}_0$ is varied, at the $68.3\%$ confidence level, which are consistent with the results of \rcite{Agrawal:2018own,Akrami:2018ylq,Raveri:2018ddi}. When $\Omega_{k,0}$ is allowed to vary, we obtain the upper bound of $\sim0.7$ on $\lambda$ (as given in \cref{fig:mcmc_ok} and \cref{tab:MCMC_bf}), independently of whether $\dot{\phi}_0$ is varied or fixed to zero, even though the contours and the one-dimensional PDFs in the two cases are slightly different. Note, however, that it is important to vary $\dot{\phi}_0$ in the case of free $\Omega_{k,0}$ for full exploration of the phase space of the quintessence model---there is no {\it a priori} reason for setting it to zero. It is, therefore, true that allowing $\Omega_{k,0}$ to vary pushes the upper bound on $\lambda$ to larger values, but the improvement is tiny and the favored values of $\lambda$ are far below the $\lambda = \sqrt{2}$ limit of the statement \eqref{dSConj}. This means that the new fixed point in the phase space of the model which arises as a result of $\Omega_k$ taking positive values, i.e., the Universe having a negative curvature, does not help with making $\lambda>\sqrt{2}$ consistent with observations---this is what we expected from the qualitative analyses of previous sections. It is, however, interesting to note that when $\Omega_{k,0}$ is allowed to vary, cosmological data seem to prefer nonzero curvature and, therefore, indicate a slight preference for an open universe, although the preference is only slightly more than $2\sigma$. We do not perform a model comparison here, but it is likely that the slightly larger upper bound on $\lambda$ for the curvature-assisted model compared to the flat one is due to the increase in the number of free parameters (i.e., the number of degrees of freedom) in the former case. As stated above, the results of the full MCMC analysis perfectly agree with the findings of the numerical grid analysis in \cref{sec:Num_Grid_Anal}. 

\section{Conclusions}
\label{sec:Conclusions}
In this paper we have studied the question of whether a nonzero (and specifically, negative) spatial curvature would make single-field quintessence models of cosmic acceleration with steep potentials observationally viable---this has been motivated by the results of \rcite{Marconnet:2022fmx,Andriot:2023wvg}.

We have first explained the importance of scalar fields in string theory, and after reviewing the theoretical reasons for why asymptotic corners of moduli space of string theoretic EFTs are interesting, we have argued that neither a de Sitter vacuum nor a
quintessence scenario with a nearly flat potential can be realized in those corners. The reason is that all known asymptotic string theory examples seem to imply that there is a lower bound on the gradient of the potential that is too steep to provide accelerated expansion consistent with cosmological observations---this has led to a conjecture that this steepness is a universal feature of quantum gravity in the asymptotic regions of moduli space, and, therefore, one may need to focus instead on the bulk of the space. However, the authors of the recent work \cite{Marconnet:2022fmx,Andriot:2023wvg} have argued that one could obtain observationally viable accelerated expansion even for steep potentials, i.e., from a fully controlled string theoretic EFT, if the Universe is negatively curved.

We have focused on the single-field quintessence model of dark energy with the exponential potential $V(\phi)=V_0e^{-\lambda \phi}$, and, after reviewing the phase space of the model and the corresponding dynamical system, we have shown, both qualitatively and through a Bayesian comparison of the model's predictions with a set of recent cosmological data, that large values of $\lambda$, i.e., values larger than the lower bound $\sqrt{2}$ given by the strong de Sitter conjecture, are observationally excluded, even for nonzero values of the spatial curvature. We have demonstrated that through a numerical grid analysis of the phase space of the model, as well as a Bayesian analysis of the model where we have scanned the parameter space using an MCMC analysis and compared various cosmological quantities with their measured values.

The observational $1\sigma$ upper bound we have found on the quantity $\lambda$ is $\sim0.6$ (consistent with the results of \rcite{Agrawal:2018own,Akrami:2018ylq,Raveri:2018ddi}) for a flat universe, where the curvature density parameter $\Omega_k$ is fixed to zero, while we have found the $1\sigma$ upper bound of $\sim0.7$ for the case in which $\Omega_{k,0}$ (the current value of the curvature density parameter) is varied as a free parameter, i.e., the curvature of the Universe is allowed to take nonzero values. Even though the curvature-assisted model allows $\lambda$ to take larger values compared to the zero-curvature model, the increase in the allowed values of $\lambda$ is too small to make the observational constraints consistent with the theoretical lower bound of $\sqrt{2}$ on $\lambda$. We have shown that our results are practically the same whether we fix the initial velocity of the scalar field $\dot{\phi}_0$ to zero or we set it free.

These results are consistent with our investigation of the phase space of the model (also allowing for an arbitrary value of the initial velocity of the scalar field), where we have performed a numerical grid analysis and shown that even though a new fixed point exists in the phase space when $\Omega_k>0$, none of the solutions of the set of differential equations describing the dynamical system simultaneously satisfy the two necessary conditions for a viable cosmology: (i) contain an extended period of matter domination before the onset of late-time cosmic acceleration and (ii) have an effective equation of state parameter that is $\sim-0.7$ at the present time.

Although the focus of our studies has been on the late-time acceleration of the Universe, we have also argued why the curvature-assisted scalar field with a steep potential cannot describe cosmic inflation in the early Universe.

This all means that if the strong de Sitter conjecture is indeed true and a universal feature of the asymptotic regions of string theory, or more generally of any theories of quantum gravity, then there seem to be two avenues to explore if one intends to explain accelerated expansion in such theories: (i) construct accelerating solutions through the bulk of moduli space instead of the asymptotic corners of it, or (ii) construct models with more than one scalar field, which have been shown to be able to provide observationally viable acceleration even with steep potentials (see, e.g., \rcite{Achucarro:2018vey,Cicoli:2020cfj,Cicoli:2020noz,Akrami:2020zfz,Eskilt:2022zky}), although this too may be challenging in asymptotic corners of string theory due to universal features of the moduli space metric.\\

{\it Note added.} In the final stages of the present work, we became aware of \rcite{Andriot:2024jsh,Bhattacharya:2024hep}, where similar results have been presented. In particular, the authors of \rcite{Bhattacharya:2024hep} have performed a cosmological data analysis of the curvature-assisted quintessence model where they have set $\dot{\phi}_0$ to zero. Although the results of our analysis show that the constraints on the parameters of the model do not strongly depend on whether $\dot{\phi}_0$ is fixed or is allowed to vary as a free parameter, this cannot be known {\it a priori}, as $\dot{\phi}_0$ is one of the phase space variables of the model and fixing it may result in missing important regions of the phase space when $\Omega_k$ takes nonzero values. We also became aware of \rcite{Ramadan:2024kmn}, where a cosmological analysis of single-field quintessence with an exponential potential has been performed for the zero-curvature model. In both cases, our results agree with theirs where they overlap.

\acknowledgements{
We thank M.~Raveri for helpful discussions. We also thank the pizza restaurant ``Internacional cuisine'' from the south of Madrid that provided us with delicious pizzas during the ChatCQG meetings that led to this work. G.A.\ is supported by the Spanish Attraccion de Talento Contract No. 2019-T1/TIC-13177 granted by the Comunidad de Madrid. M.D.\ is supported by Grant No. FPI SEV-2016-0597-19-3 from the Spanish
National Research Agency from the Ministry of Science and Innovation. I.R. is supported by the Spanish FPI Grant No. PRE2020-094163. Y.A.\ acknowledges support by the Spanish Research Agency (Agencia Estatal de Investigaci\'on)'s Grant No. RYC2020-030193-I/AEI/10.13039/501100011033, and by the European Social Fund (Fondo Social Europeo) through the  Ram\'{o}n y Cajal program within the State Plan for Scientific and Technical Research and Innovation (Plan Estatal de Investigaci\'on Cient\'ifica y T\'ecnica y de Innovaci\'on) 2017-2020. S.N.\ acknowledges support from the research Project No. PID2021-123012NB-C43. M.M. is supported by an Atraccion del Talento Fellowship 2022-T1/TIC-23956
from Comunidad de Madrid. M.M. thanks the
KITP program ``What is String Theory?'' for providing a stimulating enviroment for discussion. This research is supported in part by the Grant No. NSF PHY-1748958 to the Kavli Institute for Theoretical Physics (KITP). M.D., I.R., and M.M. acknowledge the hospitality of the
Department of Theoretical Physics at CERN and the Department of Physics of Harvard
University during the different stages of this work. M.D. thanks the ``Geometry, Strings and the Swampland Program'' workshop organized by the Max Planck Institute for Physics and the Arnold Sommerfeld Center for Theoretical Physics for providing a fantastic environment for discussions during the early stages of this work. The authors thank the Spanish Research Agency (Agencia Estatal de Investigaci\'on) through the grants IFT Centro de Excelencia Severo Ochoa CEX2020-001007-S and PID2021-123017NB-I00, funded by MCIN/AEI/10.13039/501100011033 and by ERDF A way of making Europe. The authors also acknowledge the use of the IFT Hydra cluster.
}

\bibliography{Bibliography}

\end{document}